\documentclass[twocolumn,aps]{revtex4}

\usepackage{epsf}

\begin{document}

\title{Phase Transition Of The Spin-One Square-Lattice Anisotropic 
 Heisenberg Model}

\author{D.J.J. Farnell, K.A. Gernoth, and R.F. Bishop}

\affiliation{Department of Physics, University of Manchester Institute of
  Science and Technology (UMIST), P O Box 88, Manchester M60 1QD, United 
  Kingdom}

\date{\today}

\begin{abstract}
The coupled cluster method (CCM) is applied
to the spin-one anisotropic Heisenberg antiferromagnet (HAF) 
on the square lattice at zero temperature using a 
new high-order CCM ground-state formalism for general quantum 
spin number ($s \ge 1/2$). 
The results presented constitute the
first such application of this new formalism, and they are shown 
to be among the most accurate results 
for the ground-state energy and sublattice magnetisation of this 
model as yet determined. 
We ``track'' the solution to the CCM equations at a given level 
of approximation with respect to the anisotropy 
parameter, $\Delta$, from the trivial Ising limit ($\Delta 
\rightarrow \infty$) down to a critical value $\Delta_c$, 
at which point this solution terminates. 
This behaviour is associated with a phase transition in
the system, and hence a primary result of these 
high-order CCM calculations is that they provide an accurate 
and unbiased (i.e., {\it ab initio}) estimation of the 
position of the quantum phase transition point as a 
function of the anisotropy parameter. 
Our result is, namely, that this point occurs at (or slightly below) 
the isotropic Heisenberg point at $\Delta=1$, for this model.

\end{abstract}

\maketitle

The area of quantum magnetic insulating systems at zero temperature has
become increasingly well understood for one-dimensional (or 
quasi-one-dimensional) lattices via the existence of well-known 
exact solutions such as the Bethe ansatz (BA)  \cite{ba1,ba2} and also via
more recent density matrix renormalisation group (DMRG)
calculations  \cite{DMRG1,DMRG2,DMRG3}. Similarly, zero-temperature 
quantum Monte Carlo (QMC) calculations  \cite{qmc3,qmc4} have 
been shown to yield very accurate results for spin-half, two-dimensional
systems. QMC techniques, however, are severely limited in their range of 
application by the presence of the infamous ``sign problem,''
although we note that for non-frustrated spin systems it is 
often possible to determine a ``sign rule''  \cite{sign_rules1,sign_rules2} 
which completely circumvents this minus-sign problem.
By contrast, the technique of quantum many-body 
theory known as the coupled cluster method (CCM) 
is neither limited by the presence of frustration nor by 
the spatial dimensionality of the lattice 
\cite{ccm2,ccm7,ccm4,ccm5,ccmextra,ccm6}. Indeed, a great strength of 
this method is that it is able to determine
with great accuracy the positions of quantum phase transition 
points of quantum systems within an {\it ab initio}, and thus
essentially unbiased, framework. These quantum phase transitions 
typically arise as some parameter within the Hamiltonian 
is varied, thus driving the system from one phase into 
another.

The behaviour of the quantum Heisenberg antiferromagnet (HAF) on the linear 
chain is highly dependent on whether the quantum spin number, 
$s$, has an integer or a half-odd-integer value. Indeed, Haldane 
\cite{haldane} first predicted that the spin-one isotropic HAF
would contain an excitation energy gap, and this prediction has 
subsequently and conclusively been shown numerically to be correct 
by exact diagonalisations of short chains \cite{parkinson} and 
by more recent and extremely accurate DMRG calculations \cite{DMRG3}. 
We note that this is in stark contrast to the 
exact BA solution  \cite{ba1,ba2} of the spin-half isotropic HAF for the 
linear chain, which contains no such gap. However, such exact 
diagonalisations and DMRG calculations are considerably more difficult 
for systems of higher spatial dimensionality. In this 
article, we focus on the specific case of the spin-one 
anisotropic HAF on the square lattice at zero temperature 
using the CCM, especially in relation to its quantum phase
transition, by applying a high-order CCM formalism for
general quantum spin number for the first time.

The spin-one anisotropic HAF is given by, 
\begin{equation}
H = \sum_{\langle i , j \rangle} \biggl \{ s_i^x s_j^x + 
s_i^y s_j^y + \Delta s_i^z s_j^z \biggr \} ~~ ,
\label{eq1}
\end{equation}
where the symbol $\langle i , j \rangle$ indicates 
nearest-neighbour bonds on the square lattice,
and where each bond is counted once and once only. 
For $\Delta \stackrel{>}{_{\sim}} 1$, the ground state of this model contains 
non-zero N\'eel-like order. The precise value of the 
phase transition point, at which this ordering breaks down 
is not exactly known although it is believed 
to be at (or near to) the Heisenberg point at $\Delta=1$. 
Very accurate results for the values of various ground-state 
properties of this model have been obtained via spin-wave theory 
(SWT) \cite{Hamer} 
and cumulant series expansions \cite{Zheng}. Both sets of results 
indicate that approximately $80\%$ of the classical ordering 
remains for the spin-one, square-lattice HAF model at $\Delta=1$. 
We note however that cumulant series expansions make an explicit 
assumption that the position of the phase transition point 
is exactly at $\Delta=1$ in order to perform Pad\'e resummations 
of the otherwise divergent perturbation series. Also, conventional 
SWT essentially appears to ``build in'' a phase transition point 
at $\Delta=1$ for any lattice, at which point the excitation 
spectrum becomes soft. This is, of course, now known to be incorrect 
 \cite{DMRG3,haldane,parkinson} for the spin-one linear-chain 
HAF model, as noted above.

\begin{figure}
\epsfxsize=8cm
\centerline{\epsffile{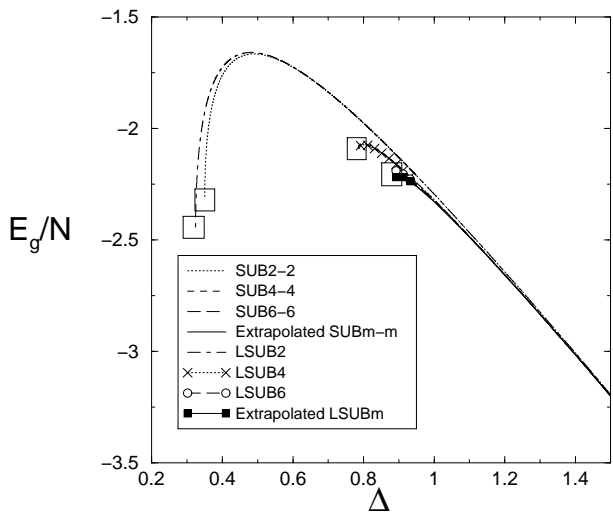}}
\vspace{0.4cm}
\caption{CCM results for the ground-state energy per spin of the 
spin-one anisotropic Heisenberg antiferromagnet on the square lattice
using the LSUB$m$ and SUB$m$-$m$ approximations with $m=\{2,4,6\}$. 
The boxes indicate the CCM critical points.}
\label{fig1}
\end{figure}

We now briefly describe the general CCM formalism, and the 
interested reader is referred to Refs. 
\cite{ccm2,ccm7,ccm4,ccm5,ccmextra,ccm6} for further details. 
The exact ket and bra ground-state energy 
eigenvectors, $|\Psi\rangle$ and $\langle\tilde{\Psi}|$, of a 
many-body system described by a Hamiltonian $H$
are parametrised within the single-reference CCM as follows:   
\begin{eqnarray} 
|\Psi\rangle = {\rm e}^S |\Phi\rangle \; &;&  
\;\;\; S=\sum_{I \neq 0} {\cal S}_I C_I^{+}  \nonumber \; , \\ 
\langle\tilde{\Psi}| = \langle\Phi| \tilde{S} {\rm e}^{-S} \; &;& 
\;\;\; \tilde{S} =1 + \sum_{I \neq 0} \tilde{{\cal S}}_I C_I^{-} \; .  
\label{ccm_eq2} 
\end{eqnarray} 
The single model or reference state $|\Phi\rangle$ is required to have the 
property of being a cyclic vector with respect to two well-defined Abelian 
subalgebras of {\it multi-configurational} creation operators $\{C_I^{+}\}$ 
and their Hermitian-adjoint destruction counterparts $\{ C_I^{-} \equiv 
(C_I^{+})^\dagger \}$. Thus, $|\Phi\rangle$ plays the role of a vacuum 
state with respect to a suitable set of (mutually commuting) many-body 
creation operators $\{C_I^{+}\}$. Note that $C_I^{-} |\Phi\rangle = 0,  
~ \forall ~ I \neq 0$, and that $C_0^{-} \equiv 1$, the identity operator.

\begin{figure}
\epsfxsize=7.3cm
\centerline{\epsffile{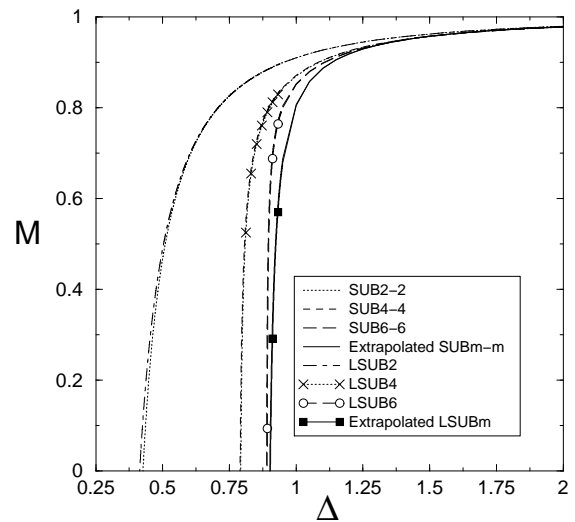}}
\vspace{0.4cm}
\caption{CCM results for the sublattice magnetisation of the 
spin-one anisotropic Heisenberg antiferromagnet on the square lattice
using the LSUB$m$ and SUB$m$-$m$ approximations with $m=\{2,4,6\}$.}
\label{fig2}
\end{figure}

We note that the exponentiated form of the ground-state CCM 
parametrisation of Eq. (\ref{ccm_eq2}) ensures the correct counting of 
the {\it independent} and excited correlated 
many-body clusters with respect to $|\Phi\rangle$ which are present 
in the exact ground state $|\Psi\rangle$. It also ensures the 
exact incorporation of the Goldstone linked-cluster theorem, 
which itself guarantees the size-extensivity of all relevant 
extensive physical quantities. 

The CCM equations are defined by the following coupled 
set of equations, 
\begin{eqnarray} 
\langle\Phi|C_I^{-} {\rm e}^{-S} H {\rm e}^S|\Phi\rangle &=& 
0 ,  ~ \forall \;\; I \neq 0 \;\; ; \label{ccm_eq7} \\ 
\langle\Phi|\tilde{S} {\rm e}^{-S} [H,C_I^{+}] {\rm e}^S|\Phi\rangle 
&=& 0 , ~ \forall \;\; I \neq 0 \; . \label{ccm_eq8}
\end{eqnarray}  
We furthermore note that the ground-state energy at the stationary 
point has the simple form 
\begin{equation} 
E_g = E_g ( \{{\cal S}_I\} ) = \langle\Phi| {\rm e}^{-S} H {\rm e}^S|\Phi\rangle
\;\; . 
\label{ccm_eq9}
\end{equation}  
It is important to realize that this (bi-)variational formulation 
does {\it not} lead to an upper bound for $E_g$ when the summations for 
$S$ and $\tilde{S}$ in Eq. (\ref{ccm_eq2}) are truncated, due to the lack of 
exact Hermiticity when such approximations are made. However, it is clear 
that the important Hellmann-Feynman theorem {\it is} still preserved in all 
such approximations. 

The CCM formalism is exact in the limit of inclusion of
all possible multi-spin cluster correlations for 
$S$ and $\tilde S$, although in any real application 
this is usually impossible to achieve. It is therefore 
necessary to utilise various approximation schemes 
within $S$ and $\tilde{S}$. The three most commonly 
employed schemes previously utilised have been: 
(1) the SUB$n$ scheme, in which all correlations 
involving only $n$ or fewer spins are retained, but no
further restriction is made concerning their spatial 
separation on the lattice; (2) the SUB$n$-$m$  
sub-approximation, in which all SUB$n$ correlations 
spanning a range of no more than $m$ adjacent lattice 
sites are retained; and (3) the localised LSUB$m$ scheme, 
in which all multi-spin correlations over distinct 
locales on the lattice defined by $m$ or fewer contiguous 
sites are retained. 
We note that in order to carry out such LSUB$m$ or SUB$n$-$m$
calculations to high order in the truncation indexes 
$n$ and $m$ we must rely on computer-algebra 
techniques in order to generate the corresponding
sets of coupled equations.  
These computational techniques are based on 
the fact that the similarity transformed Hamiltonian
in Eqs. (\ref{ccm_eq7}-\ref{ccm_eq9}) may be written {\it purely} 
in terms of creation operators acting on the model state,
so that evaluation of Eq. (\ref{ccm_eq7}) becomes purely
a matter of pattern matching. The interested reader is referred
to Refs. \cite{ccm4} and \cite{ccm6} for a full explanation of how this is
achieved for spin-half models. We note that 
the generalisation of this procedure to the case 
of arbitrary quantum spin number, $s$, is relatively 
straightforward, although detailed, and will be described
fully elsewhere \cite{ccm8}.

\begin{table}
\caption{CCM results for the ground state of the spin-one 
Heisenberg antiferromagnet at $\Delta =1$ on the square lattice using 
the LSUB$m$ approximation with $m=\{2,4,6\}$. Values for the 
CCM critical points, $\Delta_c$, of the anisotropic model 
as a function of the anisotropy, $\Delta$, are also presented.
Note that $N_F$ indicates the number of fundamental clusters
at each level of approximation.}
\begin{center}
\begin{tabular}{|@{~}c@{~}|@{~}c@{~}|@{~}c@{~}|@{~}c@{~}|@{~}c@{~}|}  \hline 
            &$N_F$  &$E_g/N$       &$M$       &$\Delta_c$    \\ \hline\hline
LSUB2       &2      &$-$2.295322   &0.909747  &0.3240        \\ \hline
SUB2\footnote{ See Ref.  \cite{ccm7}}        
            &--     &$-$2.302148   &0.8961    &0.9109        \\ \hline
LSUB4       &30     &$-$2.320278   &0.869875  &0.7867        \\ \hline
LSUB6       &1001   &$-$2.325196   &0.851007  &0.8899        \\ \hline
LSUB$\infty$ &--    &$-$2.3292     &0.8049    &0.98         \\ \hline
SWT\footnote{ See Ref.  \cite{Hamer}}          
            &--     &$-$2.3282     &0.8043    &--            \\ \hline
Series Expansions\footnote{ See Ref.  \cite{Zheng}}
            &--     &$-$2.3279(2)  &0.8039(4) &--            \\ \hline
\end{tabular}
\end{center}
\label{tab1}
\end{table}

We now apply the CCM formalism outlined
above to the specific case of the anisotropic HAF, 
and we choose the N\'eel state, in which the spins lie 
along the $z$-axis, to be the model state. Furthermore, 
we perform a rotation of the local axes of the up-pointing 
spins by 180$^\circ $ about the $y$ axis, so
that spins on both sublattices may be treated
equivalently. The (canonical) transformation is described by,
\begin{equation}
s^x \; \rightarrow \; -s^x, \; s^y \; \rightarrow \;  s^y, \;
s^z \; \rightarrow \; -s^z  \; .
\end{equation}
The model state now appears $mathematically$ to consist
of purely down-pointing spins in these rotated local
axes. In terms of the spin raising and lowering operators
$s_k^{\pm} \equiv s_k^x \pm {\rm i} s_k^y$ the Hamiltonian 
may be written in these local axes as,
\begin{equation}
H = -\frac 14 \sum_{\langle i, j \rangle}  \biggl \{ 
s_i^+ s_{j}^+ + s_i^-s_{j}^- + 2 \Delta s_i^z s_{j}^z   
\biggr \} ~~ ,
\label{eq:newH}
\end{equation}
where $i$ and $j$ runs over all nearest neighbours, although
each nearest-neighbour bond is counted once and once only. 
Furthermore, the sublattice magnetisation, $M$, (after rotation
of the local spin axes) is given by,  
\begin{equation}
M = -\frac 1{N} \sum_{i=1}^{N} s_{i}^z  ~~ .
\label{ccm_j_j'_3}
\end{equation}
In the limit $\Delta \rightarrow \infty$ we note that
the model state is the exact ground eigenstate of the 
Hamiltonian of Eq. (\ref{eq:newH}). Hence, all of the CCM 
correlation coefficients are zero at this point. We may track
this solution for decreasing values of $\Delta$ until we 
reach a {\it critical point}, $\Delta_c$, at which point  
the real solution to our CCM equations for the LSUB$m$ and SUB$m$-$m$ 
approximation schemes terminates. This is associated 
with a phase transition in the real system \cite{ccm4,ccm5,ccmextra,ccm6}. 
Similar behaviour was also seen for this model for the CCM SUB2 
approximation for this model (see Ref.  \cite{ccm7}), and once 
more this was associated with a phase transition of the 
real system. 

\begin{table}
\caption{CCM results for the ground state of the spin-one 
Heisenberg antiferromagnet at $\Delta =1$ on the square lattice using 
the SUB$m$-$m$ approximation with $m=\{2,4,6\}$. Values for the 
CCM critical points, $\Delta_c$, of the anisotropic model 
as a function of the anisotropy, $\Delta$, are also presented.
Note that $N_F$ indicates the number of fundamental clusters
at each level of approximation.} 
\begin{center}
\begin{tabular}{|@{~}c@{~}|@{~}c@{~}|@{~}c@{~}|@{~}c@{~}|@{~}c@{~}|}  \hline 
            &$N_F$  &$E_g/N$       &$M$       &$\Delta_c$  \\ \hline\hline
SUB2-2      &1      &$-$2.295041   &0.910013  &0.3499      \\ \hline
SUB2\footnote{ See Ref.  \cite{ccm7}}        
            &--     &$-$2.302148   &0.8961    &0.9109      \\ \hline
SUB4-4      &15     &$-$2.319755   &0.871195  &0.7843      \\ \hline
SUB6-6      &375    &$-$2.324863   &0.852559  &0.8879      \\ \hline
SUB$\infty$ &--     &$-$2.3291     &0.8067    &0.98        \\ \hline
SWT\footnote{ See Ref.  \cite{Hamer}}          
            &--     &$-$2.3282    &0.8043    &--         \\ \hline
Series Expansions\footnote{ See Ref.  \cite{Zheng}}
            &--     &$-$2.3279(2)  &0.8039(4)  &--         \\ \hline
\end{tabular}
\end{center}
\label{tab2}
\end{table}

We note that the ``raw'' results for the ground-state energy,
sublattice magnetisation, and critical points, $\Delta_c$,
have been obtained using both the LSUB$m$ and SUB$m$-$m$ 
approximation schemes. For both approximation schemes
we may extrapolate these results as a function of 
$m$ in the limit $m \rightarrow \infty$ in order to obtain 
even better results. We note that SUB2-$m$ calculations 
and the full SUB2 calculation have also previously been 
performed  \cite{ccm7} for the anisotropic square-lattice
HAF with general quantum spin number. It was therefore 
possible to determine accurately the manner 
in which SUB2-$m$ results scale as a function of $m$,
namely, that as $m \rightarrow \infty$ 
the ground-state energy and the critical 
points, $E_g(m)/N$ and $\Delta_c(m)$ respectively, scale 
linearly with $1/m^2$, and the sublattice magnetisation, 
$M(m)$, scales linearly with $1/m$. By analogy, we 
utilise a similar procedure here in order to extrapolate 
raw LSUB$m$ and SUB$m$-$m$ results. However, as we may 
utilise LSUB$m$ and SUB$m$-$m$ results for $m=\{2,4,6\}$ 
only, a quadratic function is fitted to these data in 
order to obtain the best possible results. A full 
and comprehensive explanation of the extrapolation
process of CCM LSUB$m$ expectation values is 
given in Ref.  \cite{ccm6} for the spin-half {\it XXZ} model
for a variety of lattices, and the interested 
reader is referred to this article for further information.

CCM results for the position of the critical point using 
the LSUB$m$ and SUB$m$-$m$ approximation schemes are 
presented in Tables \ref{tab1} and \ref{tab2}. The 
extrapolated result of $\Delta_c = 0.98$ for both
the LSUB$m$ and for the SUB$m$-$m$ approximation 
scheme indicates that the phase transition point 
is at (or perhaps slightly below) the isotropic 
Heisenberg point ($\Delta=1$). 
The strength of the CCM is that it can provide such
an accurate value for the position of the quantum
phase transition point as a function of the anisotropy
within an {\it ab initio}, and thus fully unbiased, 
framework.

Figure \ref{fig1} and Tables \ref{tab1} and \ref{tab2} indicate 
that the results for the ground-state energy of the spin-one HAF 
converge extremely quickly with $m$ for both the LSUB$m$
and SUB$m$-$m$ schemes. Indeed, even the ``raw'' unextrapolated 
results provide excellent estimates of the ground-state energy,
although the extrapolated results of $E_g/N=-2.3292$ for the
LSUB$m$ approximation scheme and $E_g/N=-2.3291$ for
SUB$m$-$m$ approximate scheme are certainly even more accurate.
By way of comparison, we note that third-order SWT 
\cite{Hamer} gives a result of $E_g/N=-2.3282$ and cumulant 
series expansions \cite{Zheng} a result of $E_g/N=-2.3279(2)$.

CCM results for the sublattice magnetisation are found to be similarly 
well converged as a function of the truncation index 
$m$ for both the LSUB$m$ and SUB$m$-$m$ 
schemes, as indicated in Fig. \ref{fig2}. The extrapolated results of 
$M=0.8049$ and $M=0.8067$ for the LSUB$m$ and SUB$m$-$m$ schemes, 
respectively, are again in excellent agreement with the results of
third-order SWT \cite{Hamer} and cumulant series 
expansions \cite{Zheng}, which give values of $M=0.8043$ 
and $M=0.8039(4)$ respectively.

In this article it has been shown that the coupled cluster method 
(CCM) provides quantitatively accurate results 
for the ground-state properties of the spin-one square-lattice 
isotropic HAF by comparison with results of third-order spin-wave 
theory (SWT) and cumulant series expansions. Furthermore, we note 
that the results presented in this article constitute the first application 
of the CCM using a new high-order CCM ground-state formalism for 
general ($s \ge 1/2$) quantum spin number. (A fuller account of
this new high-order formalism will be published elsewhere
\cite{ccm8}.) The best estimates of the ground-state 
energy of this model were found to be $E_g/N=-2.3292$ using the 
LSUB$m$ approximation scheme and $E_g/N=-2.3291$ using the
SUB$m$-$m$ approximation scheme via heuristic extrapolation to the (exact) 
limit $m \rightarrow \infty$. The best estimates of the sublattice 
magnetisation are $M=0.8049$ using the LSUB$m$ approximation
scheme and $M=0.8067$ using the SUB$m$-$m$ approximation scheme,
again via extrapolation to the limit $m \rightarrow \infty$.
The most important result of the CCM calculations presented here 
for this model is the prediction that the phase transition 
point of the spin-one square-lattice anisotropic HAF is at (or 
slightly below) the isotropic Heisenberg point. 
A strength of the method is that this prediction may be made 
using an unbiased (i.e., {\it ab initio}) treatment.

\end{document}